\documentclass[12pt]{article}

\usepackage{epsfig, bm}

\title{\bf The Site-Diluted Ising Model in Four Dimensions}
\author{ 
{\it A. Gordillo-Guerrero$^{\,1}$,} {\it R.~Kenna$^{\,2}$} and {\it J.J. Ruiz-Lorenzo$^{\,3}$}\\~\\
$^1$ Departamento de Ingenier\'{\i}a El\'ectrica, Electr\'onica y Autom\'atica,\\
 Universidad de Extremadura,\\
Avda Universidad s/n, \\
C\'aceres, 10071 Spain. 
{}\\~\\
$^2$ Applied Mathematics Research Centre,\\
Coventry University,\\
Coventry, CV1 5FB, England
{}\\~\\
$^3$ Departamento de F\'{\i}sica,\\
 Universidad de Extremadura,\\
Avda Elvas s/n, \\
Badajoz, 06071 Spain. 
{}\\~\\
}
\begin{document}
\maketitle
{\Large
  \begin{abstract}
%
In the literature, there are five distinct, fragmented
sets of analytic predictions for the scaling behaviour at the 
phase transition in the random-site Ising model in four dimensions. 
Here, the scaling relations for logarithmic corrections are used to 
complete the scaling pictures for each set. 
A numerical approach is then used to confirm the leading scaling picture 
coming from these predictions and to discriminate between them
at the level of logarithmic corrections. 
%
                        \end{abstract} }
%
  \thispagestyle{empty}
%
%
  \newpage
%
                  \pagenumbering{arabic}

\section{Introduction}
\label{I}
\setcounter{equation}{0}
One of the major achievements of statistical physics is the fundamental  explanation 
of critical behaviour at continuous phase transitions through Wilson's renormalization group (RG) approach. 
While this has mostly provided a satisfying picture 
for over thirty years, certain types of phase transitions
have resisted full treatment. Such stubborn cases, which have been the
subject of conflicting proposals and analyses,  
include systems with in-built disorder.

The Ising model with uncorrelated, quenched random-site or random-bond 
disorder is a classic example 
of such systems and has been controversial in both two and four dimensions. 
In these dimensions,
the leading exponent $\alpha$ which characterises the specific-heat 
critical behaviour vanishes and 
no Harris prediction for the consequences of quenched disorder can be 
made~\cite{Ha74}.
The Harris criterion indicates if the specific heat of the pure system 
diverges -- i.e., if $\alpha > 0$ there -- 
then the critical exponents may change as random quenched disorder 
is added to a system. 
If  $\alpha <0$ in the pure system, then this type of disorder does not alter critical behaviour
and the critical exponents are unchanged in the random system.
In the two-dimensional case, the controversy concerns the strong 
universality hypothesis
which maintains that the leading critical exponents remain the same as in the pure case and the weak universality
hypothesis, which favours dilution-dependent leading critical exponents
(see~\cite{KeRu08} and references therein). 

Since $d=4$ marks the upper critical dimensionality of the model, 
the leading critical exponents there must be given by mean field theory
and there is no weak universality hypothesis.
However, unusual corrections to scaling characterise this model, 
and the precise nature of these corrections has been debated.
This debate motivates the work presented herein: methods similar to 
those employed in ~\cite{KeRu08},
namely a high-statistics Monte Carlo approach coupled with 
finite-size scaling (FSS), are used 
to progress our understanding of the  four-dimensional version of the 
random-site Ising model (RSIM).

While not directly experimentally accessable, the four-dimensional 
RSIM  is of interest for the
following reasons: (i) it is closely related to the experimentally 
important dipolar Ising systems in three dimensions,
(ii) it is an important testing ground for the widespread 
applicability of the RG, (iii) it presents unusual corrections
to scaling, (iv) in high energy physics, the establishment 
of a non-trivial Higgs sector~\cite{FeFro92} for the standard model requires
a non-Gaussian fixed point and a new universality class which 
may, in principle, result from site dilution and (v) it is 
the subject of at least {\emph{five}}  analytical papers which 
differ in the detail of the scaling behaviour at the phase transition.

In Section~\ref{S2}, the current status of the RSIM in four dimensions is reviewed, paying particular attention to previous  detailed
analytical predictions in the literature for its scaling behaviour. 
The scaling relations recently presented in~\cite{KeJo06a,KeJo06b}
are then used to construct full scaling descriptions  based on earlier
partial theories~\cite{Ah76,Boris,Jug,GeDe93,BaFe98}. In Section~\ref{S3} 
the theoretical finite-size scaling behaviour of the model is presented.
The details of the extensive numerical simulations are given in Section~\ref{S4}, the outcomes of which are
analysed in Section~\ref{S5}. Our conclusions are summarised in 
Section~\ref{Send}.

\section{Scaling in the RSIM in Four Dimensions}
\label{S2}
\setcounter{equation}{0}

The consensus in the literature is that the following structure characterises the scaling behaviour of the
specific heat, the susceptibility and the correlation length at the second-order phase transition in the RSIM in four dimensions 
(up to higher-order correction to scaling terms)
\cite{Ah76,Boris,Jug,GeDe93,BaFe98,HeJa06}:
\begin{eqnarray} 
C_\infty(t)    & \approx & A - B|t|^{-\alpha}    \exp{ \left( {-2  \sqrt{\frac{6}{53}|\ln{|t|}|} }\right) }     |\ln{|t|}|^{\hat{\alpha}} \,, 
 \label{C1}
\\
\chi_\infty(t) & \sim & |t|^{-\gamma}          \exp{\left({\sqrt{\frac{6}{53}|\ln{|t|}|} }\right)}         |\ln{|t|}|^{\hat{\gamma}} \,, 
 \label{chiinfty} 
\\
\xi_\infty(t) & \sim & |t|^{-\nu}          \exp{\left(\frac{1}{2}{\sqrt{\frac{6}{53}|\ln{|t|}|} }\right)}          |\ln{|t|}|^{\hat{\nu}} \,.
\label{xiinfty}
\end{eqnarray}
Here, the subscript indicates the size of 
the system,  the reduced temperature $t=(T-T_c)/T_c$
marks the distance of the temperature 
$T$ from its critical value $T_c$ and  $A$ and $B>0$ are  constants. 
The correlation function at criticality decays as~\cite{Boris,GeDe93}
\begin{equation}
{\cal{G}}_\infty(x)     =  x^{-(d-2+\eta)}       |\ln{x}|^{\hat{\eta}} \,,
\label{G}
\end{equation}
where $x$ measures distance across the lattice, the 
dimensionality of which is $d$. The correlation length 
for a system of finite linear extent $L$ also exhibits a logarithmic correction and is of the form
\begin{equation}
 \xi_L(t=0) \sim L  (\ln{L})^{\hat{q}}
\,.
\label{xiq}
\end{equation}
The leading power-law behaviour is believed to be
mean field because the fixed point is expected to be Gaussian and
\begin{equation}
 \alpha = 0\,, \quad \beta = \frac{1}{2}\,, \quad \gamma = 1\,, \quad \delta = 3\,, \quad \nu = \frac{1}{2}\,, \quad \eta = 0\,, \quad \Delta = \frac{3}{2}\,.
\label{MF}
\end{equation}
Here, $\beta$ and $\delta$ are, in standard notation, the critical
exponents for the magnetization out of field and in field, respectively
while $\Delta$ is the gap exponent characterising the Yang-Lee 
edge.  
There is no dispute in the literature regarding
these leading exponents, some of which will be re-verified in this work. 
Nor is there any dispute regarding the details of the unusual exponential 
correction terms in (\ref{C1})--(\ref{xiinfty}).
However there are at least {\emph{five}} different sets of 
predictions for the exponents of the logarithmic terms,
 which differ from their counterparts in the pure model, and 
a principle aim of this work is to investigate these predictions numerically.

Aharony used a two-loop renormalization-group analysis to derive the unusual exponential terms in (\ref{C1})--(\ref{xiinfty}), and also found~\cite{Ah76} 
\begin{equation}
 \hat{\alpha}= \frac{1}{2}\,, \quad \hat{\gamma} = 0\,, \quad \hat{\nu}=0\,.
\label{Ah76}
\end{equation}
In~\cite{Boris}, Shalaev pointed out that Aharony's results needed 
to be refined and, by determining the beta function 
to three loops,  gave predictions for the specific heat and the susceptibility which differ from those in~\cite{Ah76}
in the slowly varying multiplicative logarithmic factors: 
\begin{equation}
\hat{\alpha}= 1.2368\,, \quad \hat{\gamma} = -0.3684\,, \quad \hat{\eta}=0094\,.
\label{Borisalpha}
\end{equation}
Jug studied the $\alpha = 0$ line  of $n$-component spin models 
in $(n,d)$ space where $d$ is the system's dimensionality, 
and thereby worked out the logarithmic corrections
for the $d=4$ $n$-vector model~\cite{Jug}. 
For the case at hand ($n=1$), he obtained
\begin{equation}
\hat{\alpha}= 1/2\,, \quad \hat{\gamma} = 1/212 \approx 0.0047\,.
\label{Jugalpha}
\end{equation}
In~\cite{GeDe93}, Geldart and De'Bell confirmed that to obtain the correct powers of $|\ln{|t|}|$ the beta function has 
to be calculated to three loops, but the results of~\cite{GeDe93}
differ from those of 
\cite{Boris} in the powers of the logarithms
which appear in the specific heat and in the correlation function:
\begin{equation}
 \hat{\alpha}\approx 1.2463, \quad \hat{\gamma} \approx -0.3684, \quad \hat{\eta}=\frac{1}{212}=0.0047.
\label{GaDe93}
\end{equation}
Finally Ballesteros et al.~\cite{BaFe98} extended and corrected Aharony's computation to give the correction exponents:
\begin{equation}
 \hat{\alpha}= \frac{1}{2}, \quad \hat{\gamma} = \frac{1}{106} \approx 0.0094, \quad \hat{\nu}=0, \quad \hat{q}=\frac{1}{8}.
\label{BaFe98}
\end{equation}

So the detailed analytic scaling predictions of at least five groups of authors clash, and a number of questions arise:
(i) are the predictions from within each author set self-consistent?
(ii) what are the full set of predictions (i.e., extended to all 
observables) coming from each set?
(iii) can a simulational approach yield numerical support for the shift in the
correction terms from their counterparts in the pure model, and, further,
(iv) can such a computational approach support one or other of these five
different sets of analytic predictions?
Here the scaling relation for logarithmic corrections developed in~\cite{KeJo06a,KeJo06b} are used
to accomplish (ii) and it is shown that the answers to questions (i) and (iii) 
and to some extent (iv) are in the affirmative. 
In particular, numerical support for the broad scenarios presented in~\cite{Ah76,Jug,BaFe98} is presented.

Modification of the self-consistent scaling theory
for logarithmic corrections of~\cite{KeJo06a,KeJo06b}, 
to incorporate the exponential terms, leads to the following forms 
for the  behaviour of the magnetization in the 4D RSIM:
\begin{eqnarray}
m_\infty(t)   & = & t^{\beta}        \exp{\left({-\frac{1}{2}\sqrt{\frac{6}{53}|\ln{|t|}|} }\right)}       |\ln{t}|^{\hat{\beta}} \,, 
\label{m1}
\\
m_\infty(h) & = & h^{\frac{1}{\delta}}  |\ln{h}|^{\hat{\delta}} \,.
\label{m2}
\end{eqnarray}
From Eq.(15) of~\cite{KeJo06a}, we also write for the scaling 
of the Yang-Lee edge
\begin{equation}
 r_{\rm{YL}}(t) \sim t^{\Delta} \exp{\left({-\frac{3}{2}\sqrt{\frac{6}{53}|\ln{|t|}|}}\right)} |\ln{t}|^{\hat{\Delta}}
\,.
\label{r}
\end{equation}
The scaling relations for logarithmic corrections in this
4D model are~\cite{KeJo06a,KeJo06b}\footnote{
The relation (\ref{J1}) is modified to read $ \hat{\alpha}  = 1+  d \hat{q} -  d \hat{\nu}$ when $\alpha = 0$
and when the impact angle of Fisher zeros onto the real axis is any value other than $\pi/4$, which is not expected to be the case in this 4D model~\cite{KeJo06b}.}
\begin{eqnarray}
 \hat{\alpha} & = & d \hat{q} -  d \hat{\nu}  \,,
\label{J1} \\
 2 \hat{\beta} - \hat{\gamma} & = &   d \hat{q} -  d \hat{\nu}  \,, 
\label{R1}
\\
 \hat{\beta} (\delta - 1) & = &  \delta \hat{\delta} - \hat{\gamma} \,, 
\label{G1}
\\
\hat{\eta} & = & \hat{\gamma} - \hat{\nu} (2 - \eta )\,,
\label{F1}
\\
 \hat{\Delta} & = & \hat{\beta} - \hat{\gamma}\,.
\label{A1}
\end{eqnarray}

These scaling relations are now used  to generate a complete scaling picture from the 
fragments available in the literature~\cite{Ah76,Boris,Jug,GeDe93,BaFe98}. 
This complete picture is given in Table~\ref{exp_analytic}, where the exponents of the logarithmic correction terms are listed.
Values for the exponents in boldface come directly from the reference concerned and the remaining values
are consequences of the scaling relations  (\ref{J1})--(\ref{A1}) .
\begin{table}
\caption{Theoretical predictions for the exponents of the logarithmic corrections to scaling for the pure Ising model in 
four dimensions and for its random-site counterpart. 
The latter exponents are listed in boldface 
if they come directly from the cited literature. 
The remaining values are extended from those of the literature using the
scaling relations (\ref{J1})--(\ref{A1}). 
}  
\begin{center}
\begin{tabular}{|r|l|l|l|l|l|l|} \hline \hline
Log        & Pure   model                     &  Aharony ~\cite{Ah76}    & Shalaev~\cite{Boris} & Jug~\cite{Jug} & Geldart                               & Ballesteros   \\
exp           & \cite{BaFe98,KeLa94}   &                                      &                                  &                        &  \& De'Bell~\cite{GeDe93}  &  et al ~\cite{BaFe98}   \\
\hline
$\hat{\alpha}$    & 1/3                           & {\bf{0.5}}                        &   \,{\bf{1.237}}             &   {\bf{0.5}}       &  \, {\bf{1.246}}                        &  {\bf{0.5}}      \\
$\hat{\beta}$      &  1/3                          &  0.25                              &   \,0.434                      &      0.252         &  \, 0.439                                 &  0.255           \\
$\hat{\gamma}$ & 1/3                           & {\bf{0}}                           &  {\bf{-0.368}}             &  {\bf{0.005}}    &   {\bf{-0.368}}                       &  {\bf{0.009}}  \\
$\hat{\delta}$     &  1/3                          &  0.167                            &  \,0.167                       &     0.170          &  \,  0.170                                & 0.173            \\
$\hat{\nu}$         & 1/6                           &  {\bf{0}}                          &  -0.189                      &                        &   -0.187                                & {\bf{0}}          \\
$\hat{\eta}$        & 0                               & 0                                   &  \,{\bf{0.009}}              &                        &  \, {\bf{0.005}}                         & 0.009            \\
$\hat{q}$            & 1/4                           &  0.125                            &   \,0.120                      &                        &   \,  0.125                            & {\bf{0.125}}   \\
$\hat{\Delta}$     & 0                             &  0.25                              &   \,0.803                       &    0.248            &  \, 0.807                                  & 0.245            \\
\hline \hline
\end{tabular}
\label{exp_analytic}
\end{center}
\end{table}
Each of the five papers~\cite{Ah76,Boris,Jug,GeDe93,BaFe98} is self-consistent in that the exponents
given within do not violate logarithmic scaling relations. 
However, there are clear discrepancies {\emph{between}} 
each of the five papers.

The presence of the special exponential corrections has recently 
been verified by Hellmund and Janke in the case of 
the susceptibility~\cite{HeJa06}. 
These exponential terms mask the purely logarithmic corrections, 
so in order to detect and measure the latter
one needs to cancel the former. 
Certain combinations of thermodynamic functions achieve this,
but it turns out that FSS does this also.
FSS therefore offers an ideal method to determine the exponents of the logarithmic corrections numerically~\cite{KeRu08}.

\section{Finite-Size Scaling}
\label{S3}
\setcounter{equation}{0}

Fixing the ratio of  $\xi_\infty(t)$ in (\ref{xiinfty}) and $\xi_L(0)$ in (\ref{xiq}) to $x$, one has
\begin{equation}
 t^{-\nu}        \exp{\left(\frac{1}{2}{\sqrt{\frac{6}{53}|\ln{|t|}|} }\right)}    |\ln{|t|}|^{\hat{\nu}} 
 =
 x 
 L (\ln{L})^{\hat{q}}
\,.
\label{star}
\end{equation}
Taking logarithms of both sides, one obtains 
\begin{equation}
| \ln{|t|}|  \approx \frac{1}{\nu} {\ln{L}}\,,
\label{lnLt}
\end{equation}
which re-inserted into (\ref{star}) gives
\begin{eqnarray}
 t & \sim & 
 L^{-\frac{1}{\nu}} 
 \left({ \ln{L} }\right)^{ \frac{ \hat{\nu}-\hat{q} }{\nu} }
 \exp{\left({ \frac{1}{2\nu} \sqrt{\frac{6}{53} \frac{1}{\nu} \ln{L}} }\right)}
 \left\{{
    1 + {\cal{O}} 
                     \left({ 
                              \frac{1}{\sqrt{\ln{L}}} 
                       }\right)
 }\right\}
\\
 & \sim & 
L^{-2} 
 \left({ \ln{L} }\right)^{ -\frac{\hat{\alpha}}{2} }
 \exp{\left({ \sqrt{\frac{12}{53} \ln{L}} }\right)}
 \left\{{
    1 + {\cal{O}} 
                     \left({ 
                              \frac{1}{\sqrt{\ln{L}}} 
                       }\right)
  }\right\}\,,
\label{FSSpres}
\end{eqnarray}
having used the mean-field value (\ref{MF}) for the leading exponent 
$\nu$ and the logarithmic scaling relation (\ref{J1}).
If $\hat{\alpha}= 1/2$, this recovers a result in~\cite{BaFe98} for the FSS of the pseudocritical point.

Inserting (\ref{FSSpres}) into (\ref{xiinfty}) recovers (\ref{xiq}), as it should.
The FSS of the remaining functions are determined by inserting (\ref{FSSpres}) into  (\ref{C1}) to (\ref{xiinfty}) and (\ref{m1}) to (\ref{r}).
One finds 
\begin{equation}
C_L(0)    \approx A           - B^\prime   L^{\frac{\alpha}{\nu}}
   \exp{\left({ -\left({2+\frac{\alpha}{2\nu}}\right)\sqrt{\frac{6}{53\nu} \ln{L}} }\right)}    
(\ln{L})^{\hat{\alpha} + \frac{\alpha }{\nu}(\hat{q}-\hat{\nu})} \,, 
\label{C3long}
\end{equation}
where $B^\prime \propto B$ is a positive constant ~\cite{Ah76,Boris,Jug,GeDe93,BaFe98}.
Inserting the mean-field values $\alpha = 0$, $\nu = 1/2$, one obtains the
simpler form
\begin{equation}
C_L(0)    \approx A           - B^\prime   
\exp{\left({ -2\sqrt{\frac{12}{53} \ln{L}} }\right)}    (\ln{L})^{\hat{\alpha}} \,.
\label{C3}
\end{equation}
Similarly,  the FSS for the susceptibility is
\begin{equation}
\chi_L(0)  \sim  L^{\frac{\gamma}{\nu}} 
                        |\ln{L}|^{\hat{\gamma}-\frac{\gamma}{\nu}(\hat{\nu} - \hat{q})}
 = L^{2} 
                        |\ln{L}|^{\hat{\zeta} }
\,,
\label{chi3}
\end{equation}
where 
\begin{equation}
 \hat{\zeta}= \hat{\gamma}-2(\hat{\nu} - \hat{q})
 = \frac{1}{2} \hat{\alpha}+\hat{\gamma}\,.
\label{zeta}
\end{equation}
That for the Yang-Lee edge is 
\begin{equation}
r_1(L)  \sim   L^{-\frac{\Delta}{\nu}} 
                        |\ln{L}|^{\hat{\Delta}+\frac{\Delta}{\nu}(\hat{\nu} - \hat{q})}
=
 L^{-3} 
                        |\ln{L}|^{\hat{\rho} }
\,,
\label{r2}
\end{equation}
where 
\begin{equation}
 \hat{\rho}=\hat{\Delta}+3(\hat{\nu} - \hat{q})
 = -\frac{1}{4}\hat{\alpha}- \frac{1}{2}\hat{\gamma} \,.
\label{rho}
\end{equation}
Each of these also have sub-leading scaling corrections
of strength $\mathcal{O}(1/\sqrt{\ln{L}})$ times the lead behaviour.
One notes, however, that the unusual exponential terms, which swamp the 
logarithmic corrections in the
thermal scaling formulae (\ref{chiinfty})  and (\ref{r}), drop out 
of their FSS counterparts (\ref{chi3}) and (\ref{r2}).
These are therefore ideal quantities to study the logarithmic corrections. 
The theoretical analytical predictions of each of the five sources in the literature are now used to construct five possible FSS scenarios for 
the specific heat, the susceptibility and the Lee-Yang zeros.
While Jug did not calculate the critical correlator or
correlation length in 4D, the  FSS picture corresponding to~\cite{Jug} can still be constructed through the scaling relations for logarithmic corrections.  
The FSS scenarios are listed  in Table~\ref{FSS_analytic}.
\begin{table}
\caption{The exponents of the multiplicative logarithmic corrections 
to FSS for 
the magnetic susceptibility and for the Lee-Yang zeros
coming from the literature and compared to their equivalents in the pure case. 
The FSS exponents are 
$\hat{\zeta}$ for the susceptibility and
$ {\hat{\rho}}$ for the Yang-Lee edge.
}  
\begin{center}
\begin{tabular}{|r|l|l|l|l|l|l|} \hline \hline
Exponent                                                                   & Pure & Aharony& Shalaev & Jug  & Geldart \&                              & Ballesteros   \\
& model&   ~\cite{Ah76}                 &    ~\cite{Boris}                 &~\cite{Jug}                &   De'Bell~\cite{GeDe93}  &  et al ~\cite{BaFe98}   \\
\hline
Susceptibility  $\hat{\zeta}$                                                           &  1/2                          &  \,0.25                           &   \,0.25                      &  \,    0.255         &  \, 0.255                            &  0.259           \\
Lee-Yang zeros  $\hat{\rho}$                                                       & -1/4                           &  -0.125                         &  -0.125                      &  -0.127              &   -0.127                           &  -0.130  \\
\hline \hline
\end{tabular}
\label{FSS_analytic}
\end{center}
\end{table}

The remainder of this paper is concerned with  Tables~\ref{exp_analytic} and~\ref{FSS_analytic}. 
The primary objective is to  verify that 
the exponents for the logarithmic-correction terms in the RSIM are indeed different to those of the pure model. 
Once this is established, one would like to determine which of the five sets of analytical predictions are supported numerically.
From Table~2, it is clear that present-day numerics cannot 
be sensitive enough to distinguish between all  five scenarios for
the susceptibility or individual zeros. However, there are clear differences between the predictions coming from 
\cite{Ah76,Jug,BaFe98} and~\cite{Boris,GeDe93} for the specific heat
(Table~1)
and it will turn out that the numerical data is indeed sensitive enough to favour the former over the latter.

\section{Simulation of the RSIM in Four Dimensions at Various Dilution Levels}
\label{S4}
\setcounter{equation}{0}

\begin{figure}[!ht]
\begin{center}
\includegraphics[width=0.55\columnwidth, angle=270]{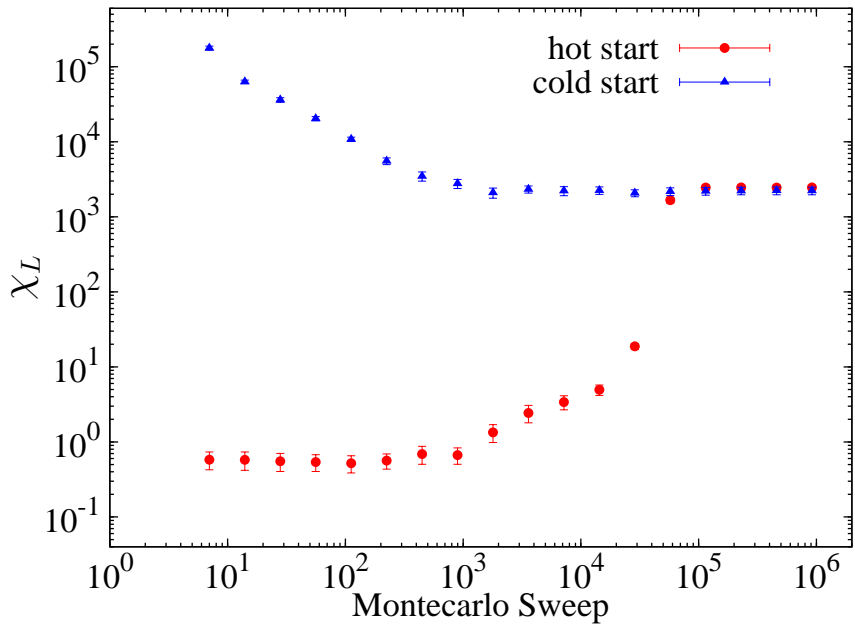}
\caption{(Color online) Averaged behaviour of the susceptibility with the 
Monte Carlo time for 20 samples at $L=32$ and $p=0.800$. 
After every Monte Carlo sweep (Wolff update)
measurements were performed. The plateau is reached more easily 
starting from a cold configuration (triangles).}
\label{termatest1}
\end{center}
\end{figure}
We have performed extensive simulations of the model for linear lattice sizes from $L=8$ to $L=48$
at dilutions $p=1$, $p=0.8$ and $p=0.5$. 
In each case, we have employed a Wolff single-cluster algorithm~\cite{Wolff89}
to update the spin variables using periodic boundary conditions.
Thermalization tests including the comparison between cold (all spins up) and 
hot (all spins random) starts have been carried out. 
We found that the plateau for the susceptibility is quickly reached by 
starting from cold configurations, see Fig.~\ref{termatest1}.
Indeed, the results for the susceptibilities from hot and cold
starts are fully compatible (and are less
than two standard deviations away from each other, even at the level
of logarithmic corrections). 
The information about the numerical details is given in Table~\ref{tablesimu}.
We have taken 1000 disorder realizations in all the cases except for
$L=48$, where only 800 samples were used. We estimate that the total 
simulation time has been equivalent to 20 years of a single node of
a Pentium Intel Core2 Quad 2.66 GHz processor.
Since our aim is to estimate the scaling of quantities right 
at the critical point, simulations  must be performed
at the critical temperature of the model. We used the estimates for 
the critical temperature  given in  ~\cite{BaFe98}. 
In terms of $\beta = 1/kT$, where $k$ is the Boltzmann constant,
these are $\beta_\mathrm{c} = 0.149695$, $\beta_\mathrm{c} =  0.188864$ and
$\beta_\mathrm{c} =  0.317368$, for $p=1$, $p=0.8$ and $p=0.5$, respectively.

\begin{figure}[!ht]
\begin{center}
\includegraphics[width=0.55\columnwidth, angle=270]{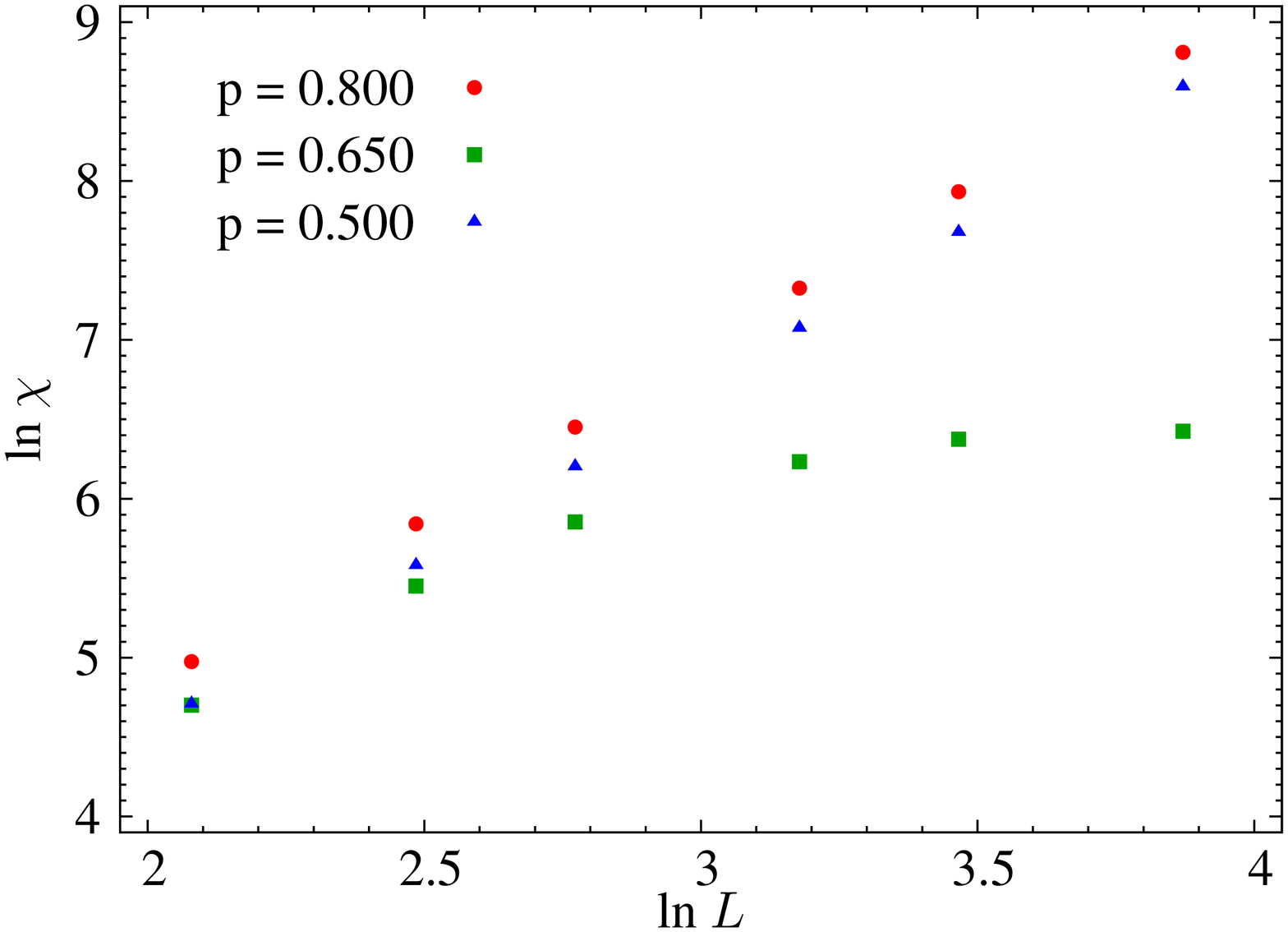}
\caption{(Color online) Comparative behaviour of the susceptibility for $p=0.650$ 
at $\beta_\mathrm{c} = 0.235049$~\cite{BaFe98}. Note
the strong deviation at this dilution from the expected leading behaviour ($\chi\sim L^2$).
The point size is in every case bigger than the corresponding error bar.
}
\label{bias_p0650}
\end{center}
\end{figure}

In addition we have simulated the dilution $p=0.650$ at $\beta_\mathrm{c} = 0.235049$~\cite{BaFe98} with
the same statistics used for the other dilutions but we have found that the behaviour of the observables
differs from that expected. In Fig.~\ref{bias_p0650}
is shown the deviation of the leading scaling behaviour of the susceptibility.
We have rechecked this point starting from different initial configurations 
and even using different random number generators. This is probably due to a biased estimation of the critical temperature in~\cite{BaFe98}. For this reason we omit $p=0.650$ from our analysis.

\begin{table}
\caption{Simulation details for each spin concentration $p$
and system size $L$.
Here, $N_\mathrm{Wolff}$ denotes the number of Wolff updates between consecutive
measures, $N_\mathrm{d}$ is the number of dropped measurements  at the
beginning of the Monte Carlo history (in units of $10^3$), 
and $N_\mathrm{m}$ is the total number of measurements performed after thermalization in units of $10^6$.
}  
\begin{center}
\begin{tabular}{|r|c|c|c|c|} \hline \hline
Spin Concentration & $L$ & $N_\mathrm{Wolff}$   & $N_\mathrm{d}$     &   $N_\mathrm{m}$ \\\hline
${\bm p}\mathbf{=1.000}$\,   &      8       &  200    &  2    &   2    \\
($\beta_\mathrm{c} = 0.149695$)	      &      12      &  400    &  8    &   4    \\
	      &      16      &  1600   &  32   &   16   \\
	      &      24      &  2000   &  128  &   20   \\
	      &      32      &  3000   &  400  &   30   \\
	      &      48      &  4000   &  1600 &   40   \\\hline
${\bm p}\mathbf{=0.800}$\,   &      8       &  100    &  1    &   0.1  \\
($\beta_\mathrm{c} =  0.188864$)	      &      12      &  200    &  4    &   0.2  \\
	      &      16      &  800    &  16   &   0.8  \\
	      &      24      &  1000   &  64   &   1    \\
	      &      32      &  1500   &  200  &   1.5  \\
	      &      48      &  2000   &  1250 &   2    \\\hline
${\bm p}\mathbf{=0.500}$\,   &      8       &  100    &  2    &   0.1  \\
($\beta_\mathrm{c} =  0.317368$)	      &      12      &  200    &  8    &   0.2  \\
	      &      16      &  800    &  32   &   0.8  \\
	      &      24      &  1000   &  128  &   1    \\
	      &      32      &  1500   &  512  &   1.5  \\
	      &      48      &  2000   &  1250 &   2    \\
\hline \hline
\end{tabular}
\end{center}
\label{tablesimu}
\end{table}

\section{Analysis of the Numerical Results}
\label{S5}
\setcounter{equation}{0}

To establish confidence in the present approach, the pure system is analysed first to test whether the method employed
successfully quantitatively identifies the logarithmic corrections which are well established there.

\subsection{The  Pure Case $p=1$}

The scaling and FSS of the pure model ($p=1$) are well understood~\cite{BaFe98,KeLa94}. 
The specific heat FSS behaviour is given by
\begin{equation}
  C_L(0)   \sim (\ln{L})^{\hat{\alpha}} \sim  (\ln{L})^{1/3}
\,,
\label{C3pure}
\end{equation}
up to additive corrections. 
Fitting to this form for $\hat{\alpha}$ over the full data set $ 8 \le L \le 48$, one finds the estimate $\hat{\alpha}= 0.42(4)$ with a goodness of fit given
by a $\chi^2/\mathrm{d.o.f.}$ (chi-squared per degree of freedom) of $1.3$. The estimate is two standard deviations away of the known value $1/3$. 
As elsewhere in this analysis, inclusion of sub-leading scaling correction terms in the fits does not ameliorate this result, which is 
similar to that reported in \cite{BaFe98}.

The FSS for the susceptibility is given in (\ref{chi3}) with $\hat{\zeta} = 1/2$. Fitting to the leading form 
\begin{equation}
 \chi_L(0) \sim  L^{\frac{\gamma}{\nu}}
\label{chifit}
\end{equation}
gives $\gamma/\nu=2.16(1)$ for $ 8 \le L \le 48$ and
 $\gamma/\nu=2.13(2)$  for $ 12 \le L \le 32$, the difference from the theoretical value $\gamma/\nu=2$ being
ascribable to the presence of the logarithmic correction term.  
Accepting this mean-field value for $\gamma/\nu$ and fitting to 
\begin{equation}
 \chi_L(0) \sim L^{2} (\ln{L})^{\hat{\zeta}}
\,,
\label{chifitlog}
\end{equation}
gives the estimate
 $\hat{\zeta} = 0.48 \pm 0.02$   in the range $8 \le L \le 48$, albeit with $\chi^2/\mathrm{d.o.f.}=4.1$

The FSS for the individual Lee-Yang zeros is given in (\ref{r2}) with $\hat{\rho} = -1/4$ in the pure case. Fitting to the leading form 
\begin{equation}
 r_j(L) \sim L^{-\frac{\Delta}{\nu}}
\label{rfit}
\end{equation}
gives $\Delta / \nu= 3.074(5)  $ for $ 8 \le L \le 48$,
the difference from the theoretical mean-field value $\Delta / \nu=3$ being
due to the corrections.  Accepting this value and fitting to 
\begin{equation}
 r_j(L) \sim L^{-3} (\ln{L})^{\hat{\rho}}
\,,
\label{rfitlog}
\end{equation}
gives  $\hat{\rho} = -0.22(2)$   in the range $8 \le L \le 48$.
This estimate is compatible with the known value $\hat{\rho}=-1/4$.
As one would expect,  the higher zeros yield less accurate estimates
(as they are further from the real simulation points) with 
 $\hat{\rho} = -0.18(3)$ ,
 $\hat{\rho} = -0.17(7)$ 
and
 $\hat{\rho} = -0.10(14)$ 
from the second, third and fourth zeros respectively. 
These estimates are listed in Table~\ref{FSS_estimates}.

Having established that the numerics gives reasonable agreement
with the  pure theory at the leading and the 
logarithmic levels, we now perform a similar analysis 
in the presence of disorder.

\subsection{The  Diluted Cases $p=0.8$ and $p=0.5$}

Since the Ansatz (\ref{C3}) for the specific heat in the disordered
systems is rather more complex than that of the pure case (\ref{C3pure}),
we begin the $p \ne 1$ analyses with the susceptibility and the 
Lee-Yang zeros. It will turn out that our analyses 
will reinforce the analytical predictions that
scaling is governed by the Gaussian fixed point and that the logarithmic corrections in the RSIM differ from 
those in the pure model. Indeed, the results for the zeros will be seen 
to be broadly compatible with the analytic
predictions contained in~\cite{Ah76,Boris,Jug,GeDe93,BaFe98}.

\begin{table}
\caption{FSS estimates for the various dilution values, using a range
of lattice sizes. The susceptibility is expected to scale as $\chi_L\sim L^2(\ln{L})^{\hat{\zeta}}$ and the Lee-Yang zeros 
as $r_j\sim L^{-3}(\ln{L})^{\hat{\rho}}$, where $\hat{\zeta} \approx 0.25$ to $0.259$ and $\hat{\rho} \approx -0.125$ to $-0.130$. (For comparison, the pure theory with $p=1$ has $\hat{\zeta} = 1/2$ and $\hat{\rho} = -1/4$.)
}  
\begin{center}
\begin{tabular}{|l|l|l|l|l|l|l|} \hline \hline
   $p$   &          & $\hat{\zeta}$&\multicolumn{4}{c|}{$\hat{\rho}$}  \\ 
\hline
         &          &             &        \multicolumn{4}{c|}{} \\ 
         & Theory ($p=1$) $\Rightarrow$& 1/2&\multicolumn{4}{c|}{ -1/4}\\ 
         & Theory ($p\ne 1$) $\Rightarrow$&$0.25$ to $0.26$&\multicolumn{4}{c|}{$-0.125$ to $-0.13$}\\
\hline
   &           &             &  $j=1$ & $j=2$  &$j=3$&$j=4$  \\ 
\hline
1  & $L=8-48$    & 0.48(2)  &-0.22(2)&-0.18(3)&-0.17(7) &-0.10(14)\\ 
\hline
0.8& $L=8-48$    & 0.39(3)  &-0.15(2)&-0.16(3)&-0.20(3) &-0.17(3) \\ 
0.8& $L=12-48$   & 0.42(4)  &-0.17(4)&-0.16(4)&-0.17(5) &-0.18(4)   \\ 
\hline
0.5& $L=8-48$    &  0.37(4) &-0.20(4)&-0.22(4)&-0.21(4) &-0.21(4) \\ 
0.5& $L=12-48$   &  0.40(6) &-0.16(5)&-0.20(5)&-0.18(5) &-0.19(5) \\ 
\hline \hline
\end{tabular}
\label{FSS_estimates}
\end{center}
\end{table}
For the weaker dilution value $p=0.8$, a fit using all lattice sizes to 
the leading form (\ref{chifit}) for the susceptibility yields
the estimate $\gamma/\nu =  2.14 \pm     0.01$.

\begin{figure}[!ht]
\begin{center}
\includegraphics[width=0.45\columnwidth, angle=270, trim=220 40 0 0, clip]{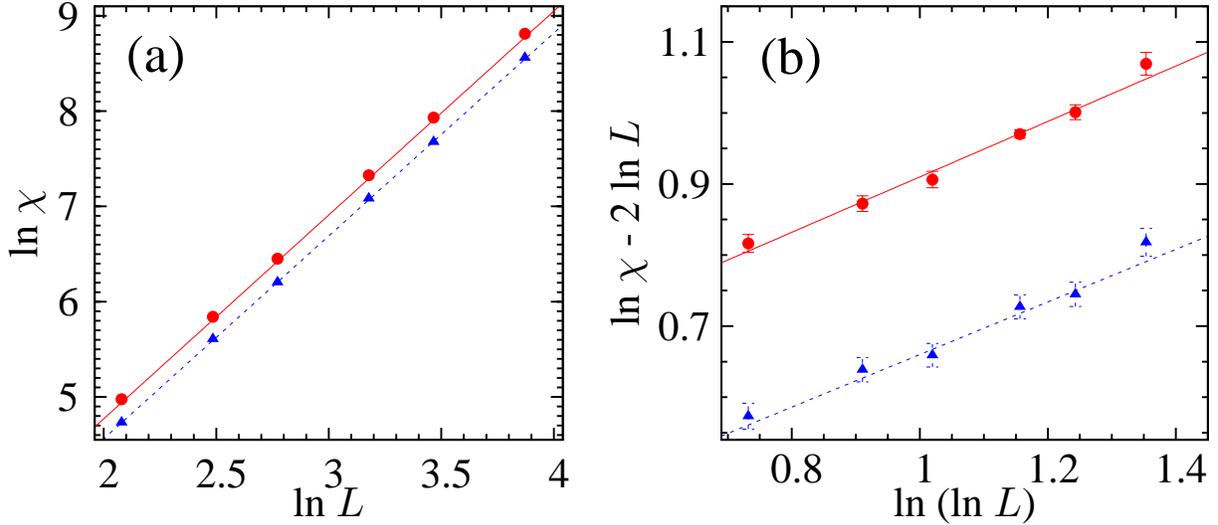}
\caption{
(Color online)
(a): FSS plot for $\chi_L$ at $p = 0.8$  (circles)
and $p=0.5$ (triangles) at the critical point. 
The slopes of the fitted solid and dashed lines give estimates for 
$\gamma / \nu$ of $2.14(1)$ and $2.13(2)$, respectively.
(b): Plot of  $\ln{\chi_L}-2 \ln{L}$ against $\ln{(\ln{L})}$ 
at $p = 0.8$  (circles) and $p=0.5$ (triangles)
giving  slopes $0.39(3)$ and $0.37(4)$, respectively, 
indicating slow crossover of  multiplicative logarithmic 
corrections from the pure case (where $\hat{\zeta} = 0.5$) to the diluted
case, where the theoretical value is $\hat{\zeta} \approx 0.13$.
}
\label{fig_susceptibility}
\end{center}
\end{figure}

Ascribing the difference from the Gaussian value $\gamma/\nu=2$ as being due
to the correction terms and, as in the pure case, and  fitting to (\ref{chifitlog}),
one finds an estimate for the correction 
exponent $\hat{\zeta} = 0.39(3)$ for $8 \le L \le 48$.
This values is between the pure value $\hat{\zeta} = 0.5$ and the
theoretical estimates for the diluted value which give $\hat{\zeta}\approx 0.25$
to $0.26$. Thus, while the FSS for the susceptibility does not capture the
theoretical estimates for the diluted case, 
the fitted values have moved away from the pure 
value and towards the lower value listed in Table~\ref{FSS_analytic}.
As elsewhere in this work, the inclusion of scaling corrections does 
not alter these results significantly.

A similar analysis for the FSS of the susceptibility at the stronger
dilution value $p=0.5$ gives similar results: the leading form
(\ref{chifit}) yields an estimate $\gamma/\nu =  2.13 \pm     0.02$
with a goodness of fit given by $\chi^2/\mathrm{d.o.f.}=0.4$.
Ascribing the difference from the mean-field value  $\gamma/\nu =  2$
as being due to the logarithmic corrections, and fitting to (\ref{chifitlog}),
one obtains the estimate $\hat{\zeta} = 0.37(4)$ for $8 \le L \le 48$.
Again this result is between the theoretical predictions for the 
pure ($\hat{\zeta} = 0.5$) and diluted ($\hat{\zeta} \approx 0.25$ to $0.26$)
cases. 
These results are summarised in Table~\ref{FSS_estimates}, together with results obtained
from the same fits with the smallest lattices removed. The best fits of the
susceptibility can be seen in Fig.~\ref{fig_susceptibility}.

Since in each of the diluted cases, 
the results for susceptibility lie between what is expected for 
the pure theory and for the diluted theories, we appeal to the Lee-Yang 
zeros, as  our collective experience indicates that they give a cleaner 
signal. 

\begin{figure}[!ht]
\begin{center}
\includegraphics[width=0.45\columnwidth, angle=270, trim=220 40 0 0, clip]{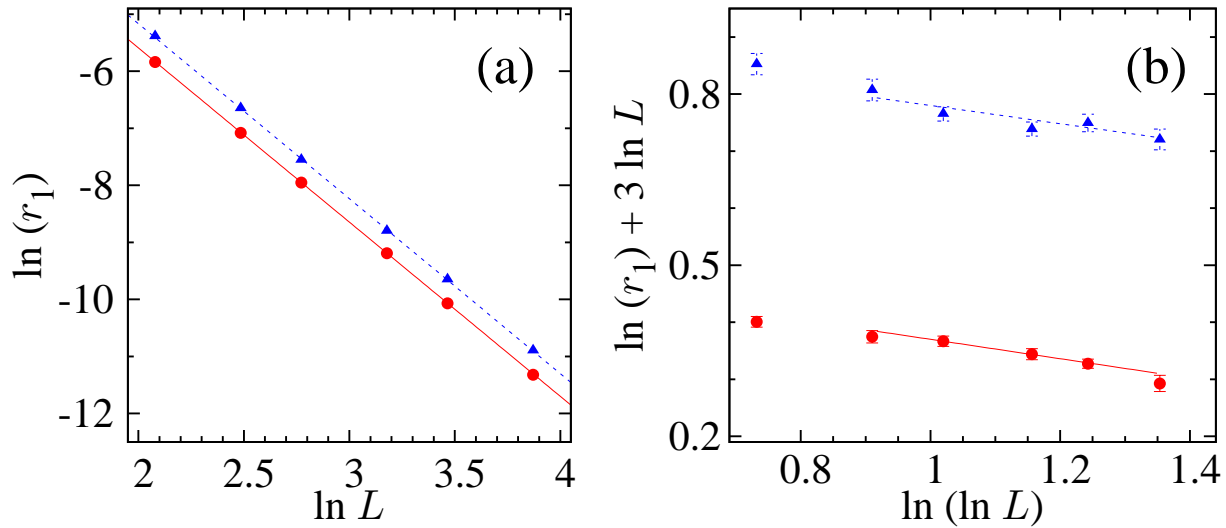}
\caption{
(Color online)
(a): FSS plot for the Yang-Lee edge at $p = 0.8$  
(circles) and $p=0.5$ (triangles). 
The slopes of the fitted solid and dashed lines give estimates for 
$\Delta / \nu$ of $3.055(4)$ and $3.07(2)$, respectively.
(b): Plot of  $\ln{r_1}+3 \ln{L}$ against $\ln{(\ln{L})}$ 
at $p = 0.8$  (circles) and $p=0.5$ (triangles).
Fits in the range $L=12$ to $L=48$ (plotted) give
 slopes $-0.17(4)$ and $-0.16(5)$, compatible with the 
predictions ranging from $\hat{\rho}\approx -0.125$ 
and $\hat{\rho}\approx -0.13$ 
in the literature.
(For comparison, in the pure model, $\hat{\rho} = -1/4$).
}
\label{fig_firstzero}
\end{center}
\end{figure}

The leading behaviour is firstly examined by fitting each of the 
first four Lee-Yang zeros to (\ref{rfit}). 
For the weaker dilution given by $p=0.8$, one obtains 
$\Delta/\nu = 3.055(8)$, $3,056(9)$, $3.069(11)$ and $3.060(10)$
from fits to the first, second, third and fourth zeros
respectively, using all lattice sizes.
The equivalent results for the stronger dilution value
$p=0.5$ are 
$\Delta/\nu = 3.068(13)$, $3,071(15)$, $3.072(12)$ and $3.071(11)$,
respectively.
All fits are of good quality with acceptable values of $\chi^2/\mathrm{d.o.f.}$,
which we refrain from detailing.
Again, these are interpreted as being supportive of the mean-field
leading behaviour $\gamma/\nu=3$ with logarithmic corrections.

The logarithmic-correction exponents are estimated by fitting to 
(\ref{rfitlog}), with the various theories indicating that 
$\hat{\rho}= -0.125$ to $-0.13$. The strongest evidence supporting this
comes, as it should, from the first zero (the 
Yang-Lee edge) for $p=0.8$, which yields the estimate $\hat{\rho} = -0.15(2)$
(with $\chi^2/\mathrm{d.o.f.}=0.6$).
As in the pure case, and as expected, estimates for $\hat{\rho}$
deteriorate as higher-index zeros are used. Dropping the smallest 
lattices from the analysis, however, leads to these estimates for  $\hat{\rho}$
more compatible with~\cite{Ah76,Boris,Jug,GeDe93,BaFe98}. These results are
summarised in Table~\ref{FSS_estimates}.

The equivalent analysis for the stronger dilution value $p=0.5$ is less clear,
with an estimate $\hat{\rho}=-0.20(4)$ coming from the first zero
when all lattices are included in the fit (with $\chi^2/\mathrm{d.o.f.}=1.1$).
Dropping the smallest  lattices, however,
gives $\hat{\rho}=-0.16(5)$ (with $\chi^2/\mathrm{d.o.f.}=0.9$),
closer to the values coming from 
~\cite{Ah76,Boris,Jug,GeDe93,BaFe98}. 
Similar results are obtained for the higher zeros and these are also 
summarised in Table~\ref{FSS_estimates}. The best fits for the first zero
are shown in Fig.~\ref{fig_firstzero}.

As a final check of the reliability of our results we have 
used the spectral energy method~\cite{FS} to reweight the data obtained at $\beta_\mathrm{c}$
to $\beta_\mathrm{c}\pm\Delta\beta_\mathrm{c}$ (taken again from~\cite{BaFe98})
obtaining that the new data sets are fully supportive of the previous results\footnote{
When doing $\beta$-extrapolations in a disordered model
one should be careful and take into account the bias induced by the finite number of measures
(see the discussions in~\cite{BaFe98,HaPa07}). We have followed the recipe provided in~\cite{BaFe98}
to perform the extrapolation to infinite number of measures per sample.
}.

Having checked that the leading FSS behaviour corresponds to that
originating in the Gaussian fixed point, and that the logarithmic
corrections to scaling are different to those in the pure model, and 
moreover (at least in the case of the Yang-Lee edge) are broadly compatible 
with the predictions from the literature~\cite{Ah76,Boris,Jug,GeDe93,BaFe98}, 
we now attempt to distinguish {\emph{between}} 
these broad predictions.  
To this end we turn to the specific heat.

\begin{figure}[!ht]
\begin{center}
\includegraphics[width=0.55\columnwidth, angle=270, trim=0 40 0 0, clip]{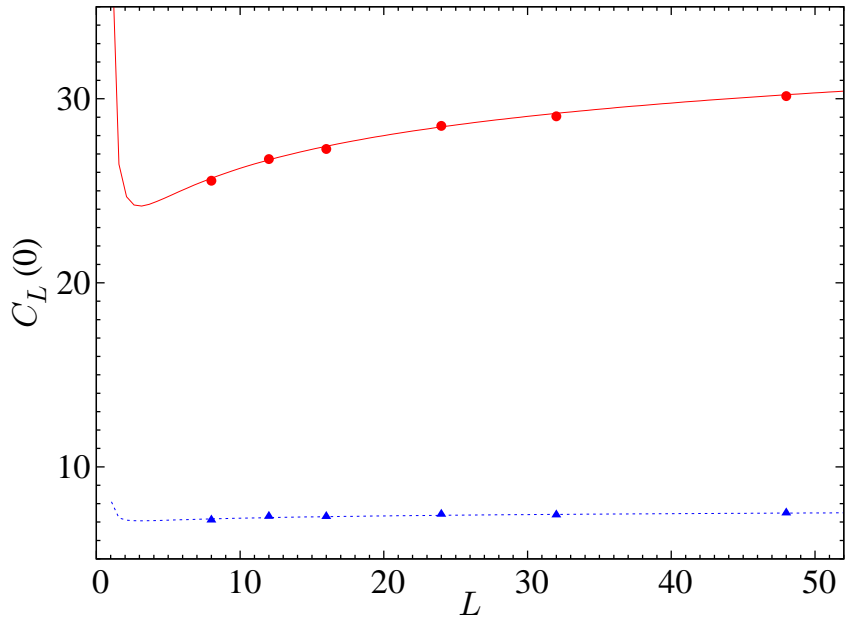}
\caption{(Color online) The specific heat for $p=0.8$ (circles) and 
$p=0.5$ (triangles). The error bars are in every case smaller than the point size. 
The solid and dashed curves are best fits to the
Ansatz~(\ref{C3}), with $\hat{\alpha}=1/2$.
}
\label{fig_cesp}
\end{center}
\end{figure}

Having established confidence that the mean-field values 
$\gamma=1$ and $\Delta=3/2$ hold in the 4D RSIM, we may use the scaling 
relation $\alpha = 2-2\Delta +\gamma $ to establish the mean-field 
value $\alpha = 0$ too. 
The Ansatz (\ref{C3}) for the specific heat may now be used.
This  contains information 
which can be used to discriminate between some of the scenarios in the
literature. From Table~\ref{exp_analytic}, there is a striking difference between the estimates for the specific-heat logarithmic-correction exponent
$\hat{\alpha}$ coming from ~\cite{Boris,GeDe93} and from~\cite{Ah76,Jug,BaFe98}. While the former have 
relatively large values of $\hat{\alpha}$, the latter agree on $\hat{\alpha} = 0.5$. 
The simulated values of the specific heat at $p=0.8$ and $p=0.5$
are plotted in Fig.~\ref{fig_cesp}.
The slope of the full specific-heat curve (\ref{C3})  is 
\begin{equation}
 \frac{d C_L}{dL} = [A-C_L(0)]\frac{\sqrt{12/53}}{L\sqrt{\ln{L}}}\left({ 1 - \frac{\hat{\alpha} \sqrt{53/12}}{\sqrt{\ln{L}}} }\right)
\,.
\label{slope}
\end{equation}
This  vanishes when 
$C_L(0) = A$ and when $\sqrt{\ln{L}} = \hat{\alpha} \sqrt{53/12}$.
The first of these is the asymptote 
$L \rightarrow \infty$, from which  $A$ can be determined
for each dilution. 
The second occurrence of zero slope is for quite small lattice sizes,
i.e., beneath lattice size $L=8$.
Therefore 
$ \hat{\alpha}
{\rm{\raisebox{-.75ex}{ {\small \shortstack{$<$ \\ $\sim$}} }}}
\sqrt{53/12}\sqrt{\ln{8}}
\approx 0.7
$, excluding the values $ \hat{\alpha}\approx 1.237$ 
and $\hat{\alpha}\approx 1.246$ given in~\cite{Boris,GeDe93}.
In fact, a best fit to the Ansatz (\ref{C3}) gives
$A=49(11)$, $B^\prime=66(22)$ and $\hat{\alpha} = 0.46(18)$ for $p=0.8$
and
$A=10(5)$, $B^\prime=7(10)$ and $\hat{\alpha} = 0.7(3)$ for $p=0.5$ with $L>8$.
Fixing $\hat{\alpha}=1/2$ in each case gives
$A=52(2)$, $B^\prime=72(5)$ for $p=0.8$
and
$A=9(3)$, $B^\prime=5(1)$  for $p=0.5$, 
and these curves are plotted along with the specific heat measurements in 
Fig.~\ref{fig_cesp}.
Fixing the correction exponent $\hat{\alpha}$ to the  
value given in~\cite{Boris,GeDe93}, on the other hand, yields a best-fit
value of $B^\prime$ which is negative in each case, contradictory to
\cite{Ah76,GeDe93}. Thus we can deem these values to be unlikely.

\section{Conclusions}
\label{Send}
\setcounter{equation}{0}

Numerical measurements of the leading critical exponents
in the 4D RSIM are presented, confirming that the phase transition 
in this model is governed by the Gaussian fixed point.
We then turn to the corrections to scaling, for which there exist 
{\emph{five distinct sets of predictions}} in the literature 
\cite{Ah76,Boris,Jug,GeDe93,BaFe98}. 
The scaling relations for logarithmic
corrections are used to render complete these sets 
and their counterparts for finite-size systems are given.

The measured values of the susceptibility FSS correction exponent 
$\hat{\zeta}$ for the site-diluted model, lie between
the known value for the pure model and the theoretical estimates
coming from ~\cite{Ah76,Boris,Jug,GeDe93,BaFe98} for the disordered system.
While this result illustrates slow crossover of the susceptibility,
the lowest lying Lee-Yang zeros give a cleaner signal and the measured
value for their logarithmic correction exponents are
 indeed compatible with the theories.

To discriminate between the five theories, the 
detailed finite-size scaling behaviour of the specific heat is also 
examined. The analysis is clearly in favour of the analytical predictions
of~\cite{Ah76,Jug,BaFe98}  over those of~\cite{Boris,GeDe93}. 
This is contrary to expectation as the former involve only two  loops 
in the perturbative RG expansion, while the latter take the expansion
to three loops in the beta function.

\vspace{1cm}
\noindent
{\bf{Acknowledgements:}} 
RK thanks the Statistical Physics Group at the 
Institut Jean Lamour, Nancy Universit{\'{e}}, France,
for hospitality during completion of this work.
We also thank Giancarlo Jug for stimulating discussions.
This work has been partially supported by MEC, contracts
FIS2007-60977 and FIS2006-08533-C03. Part of the simulations
were performed in  the BIFI and CETA-CIEMAT clusters.

%

\end{document}